\documentstyle{mn}             %% scommenta se lo vuoi "normale" %%
\newif\ifAMStwofonts
\def\n0{{n_0}}

\def\msun{${\rm M_\odot}\;$}

\def\spose#1{\hbox to 0pt{#1\hss}}
\def\lta{\mathrel{\spose{\lower 3pt\hbox{$\mathchar"218$}}
     \raise 2.0pt\hbox{$\mathchar"13C$}}}
\def\gta{\mathrel{\spose{\lower 3pt\hbox{$\mathchar"218$}}
     \raise 2.0pt\hbox{$\mathchar"13E$}}}
\ifoldfss
  \ifCUPmtlplainloaded \else
    \NewTextAlphabet{textbfit} {cmbxti10} {}
    \NewTextAlphabet{textbfss} {cmssbx10} {}
    \NewMathAlphabet{mathbfit} {cmbxti10} {} % for math mode
    \NewMathAlphabet{mathbfss} {cmssbx10} {} %  "   "    "
  \fi
  \ifAMStwofonts
    \ifCUPmtlplainloaded \else
      \NewSymbolFont{upmath} {eurm10}
      \NewSymbolFont{AMSa} {msam10}
      \NewMathSymbol{\upi}     {0}{upmath}{19}
      \NewMathSymbol{\umu}     {0}{upmath}{16}
      \NewMathSymbol{\upartial}{0}{upmath}{40}
      \NewMathSymbol{\leqslant}{3}{AMSa}{36}
      \NewMathSymbol{\geqslant}{3}{AMSa}{3E}

       \let\le=\leqslant
       
    \fi
  \fi
\fi % End of OFSS

\ifnfssone
  \newmathalphabet{\mathit}
  \addtoversion{normal}{\mathit}{cmr}{m}{it}
  \addtoversion{bold}{\mathit}{cmr}{bx}{it}
  \newmathalphabet{\mathbfit} % math mode version of \textbfit{..}
  \addtoversion{normal}{\mathbfit}{cmr}{bx}{it}
  \addtoversion{bold}{\mathbfit}{cmr}{bx}{it}
  \newmathalphabet{\mathbfss} % math mode version of \textbfss{..}
  \addtoversion{normal}{\mathbfss}{cmss}{bx}{n}
  \addtoversion{bold}{\mathbfss}{cmss}{bx}{n}
  \ifAMStwofonts
    \ifCUPmtlplainloaded \else
      \UseAMStwoboldmath
      \makeatletter
      \new@mathgroup\upmath@group
      \define@mathgroup\mv@normal\upmath@group{eur}{m}{n}
      \define@mathgroup\mv@bold\upmath@group{eur}{b}{n}
      \edef\UPM{\hexnumber\upmath@group}
      \new@mathgroup\amsa@group
      \define@mathgroup\mv@normal\amsa@group{msa}{m}{n}
      \define@mathgroup\mv@bold\amsa@group{msa}{m}{n}
      \edef\AMSa{\hexnumber\amsa@group}
      \makeatother
      \mathchardef\upi="0\UPM19
      \mathchardef\umu="0\UPM16
      \mathchardef\upartial="0\UPM40
      \mathchardef\leqslant="3\AMSa36
      \mathchardef\geqslant="3\AMSa3E

       \let\le=\leqslant

    \fi
  \fi
\fi % End of NFSS release 1

\ifnfsstwo
  \DeclareMathAlphabet{\mathbfit}{OT1}{cmr}{bx}{it}
  \SetMathAlphabet\mathbfit{bold}{OT1}{cmr}{bx}{it}
  \DeclareMathAlphabet{\mathbfss}{OT1}{cmss}{bx}{n}
  \SetMathAlphabet\mathbfss{bold}{OT1}{cmss}{bx}{n}
  \ifAMStwofonts
    \ifCUPmtlplainloaded \else
      \DeclareSymbolFont{UPM}{U}{eur}{m}{n}
      \SetSymbolFont{UPM}{bold}{U}{eur}{b}{n}
      \DeclareSymbolFont{AMSa}{U}{msa}{m}{n}
      \DeclareMathSymbol{\upi}{0}{UPM}{"19}
      \DeclareMathSymbol{\umu}{0}{UPM}{"16}
      \DeclareMathSymbol{\upartial}{0}{UPM}{"40}
      \DeclareMathSymbol{\leqslant}{3}{AMSa}{"36}
      \DeclareMathSymbol{\geqslant}{3}{AMSa}{"3E}

       \let\le=\leqslant

    \fi
  \fi
\fi % End of NFSS release 2

\ifCUPmtlplainloaded \else
  \ifAMStwofonts \else % If no AMS fonts
    \def\upi{\pi}
    \def\umu{\mu}
    \def\upartial{\partial}
  \fi
\fi

\title{Galactic Winds and Circulation of the ISM in Dwarf Galaxies}
\author[A. D'Ercole and F. Brighenti]
       {A. D'Ercole$^1$\thanks{annibale@bo.astro.it} 
       and F. Brighenti$^{2,3}$\thanks{brighenti@bo.astro.it} \\
        $^1$ Osservatorio Astronomico di Bologna, via Zamboni 33,
        Bologna 40126, Italy\\
        $^2$Dipartimento di Astronomia, Universit\`a di Bologna,
        via Zamboni 33, Bologna 40126, Italy\\
        $^3$University of California Observatories/Lick Observatory,
        Board of studies in Astronomy and Astrophysics, University of 
        California, \\ 
        Santa Cruz, CA 95064}
\date{Accepted 
      Received ;
      in original form }

\pagerange{\pageref{firstpage}--\pageref{lastpage}}
\pubyear{1994}

\begin{document}

\maketitle

\label{firstpage}

\begin{abstract}

We study, through 2D hydrodynamical simulations, the feedback of a 
starburst on the ISM of typical gas rich dwarf galaxies.
The main goal is to address the circulation of the ISM and
metals following the starburst.
We assume a single-phase rotating ISM in equilibrium in the
galactic potential generated by a stellar disk and a spherical dark halo.
The starburst is assumed to occur in a small volume in the center
of the galaxy, and it generates a mechanical power of $3.8 \times 10^{39}$
erg s$^{-1}$ or $3.8 \times 10^{40}$ erg s$^{-1}$ for 30 Myr.
We found, consistently with previous investigations, that the galactic
wind is not very effective in removing the ISM. The metal rich stellar ejecta,
instead, may be efficiently expelled from the galaxy and dispersed
in the intergalactic medium.

Moreover, we found that the central region of the galaxy is always replenished
with cold and dense gas after a few 100 Myr from the starbust,
achieving the requisite for a new star formation event in $\approx 0.5 - 1$
Gyr. The hydrodynamical evolution of galactic winds is thus consistent
with the episodic star formation regime suggested by many chemical evolution
studies.
%A possible exception to this rule happens when the galaxy is embedded
%in a tenuous hot gas confining the cold ISM.

We also discuss the X-ray emission of these galaxies and find
that the observable (emission averaged) abundance of the hot gas
underestimates the real one if thermal conduction is effective.
This could explain the very low hot gas metallicities estimated in starburst
galaxies.
%with particular
%emphasis on the hot gas abundance derivedmetallicity of the hot ISM.

\end{abstract}


\section{Introduction}

Many gas rich dwarf galaxies are known to be in a starburst phase,
or are believed to have experienced
periods of intense star formation in the past
(e.g. Gallagher \& Hunter 1984; Thuan 1991; Tosi 1998, and references therein).
These galaxies, classified as ``blue compact dwarf (BCD) galaxies'' or
``HII galaxies'',
are thus excellent laboratories to investigate
the feedback of vigorous star formation on the interstellar medium (ISM).

Massive stars inject enormous amount of energy in the ISM through
stellar winds and when they explode as type II supernovae (SNe);
the impact of such an energy input on the galactic ISM may,
in principle, be devastating.
In fact it is often found that the total energy
released during a starburst is greater that the gas
binding energy. Yet many dwarf galaxies in a post-starburst phase are still
gas rich.
As Skillman \& Bender (1995) pointed out, observational evidence (e.g. Marlowe
et al. 1995; Martin 1996) are still insufficient to substantiate
a disruptive impact of galactic winds on the ISM.
Clearly, simple energetic considerations do not catch
the essential nature of the feedback process, and detailed, time
dependent hydrodynamical models are needed.

Galactic winds are thought to have a key role in the formation
and evolution 
of dwarf galaxies (Dekel \& Silk 1986, Babul \& Rees 1992, Matteucci
\& Chiosi 1983). 
In general, understanding the physics
of the feedback of massive stars
on the ISM is a key problem in cosmological
theories of galaxy formation (Yepes et al. 1997, Cole et al. 1994).
Gas outflows from dwarf galaxies are also suggested to be an
important factor for the production and enrichment of the intergalactic
medium (Trentham 1994). However, persuasive arguments against this conclusion
are given by Gibson \& Matteucci (1997), and the origin of metals
in clusters of galaxies is still a matter of debate (Brighenti \& Mathews
1998).

The fate of the (metal rich) material ejected by massive
stars is of crucial importance in understanting the chemical evolution
of these galaxies (Tosi 1998), in particular the low $\alpha$-elements
abundance and the `strange' values of (He/H) and (N/O) vs. (O/H).
These problems have been encompassed invoking a `differential ejection',
in which the enriched gas lost by massive stars escapes from the galaxy
as galactic wind, while some (or most) of the original ISM is
unaffected.

Recent hydrodynamical simulations
have verified that, under many circumstances, galactic winds
are able to eject most of the metal rich gas, preserving a significant
fraction of the original ISM (MacLow \& Ferrara 1998, hereafter MF; De Young \&
Heckman 1994; De Young \& Gallagher 1991). Silich \& Tenorio-Tagle
(1998) and Tenorio-Tagle (1996), instead, found that even the metal-rich
material is hardly lost from galaxies, since it is at first trapped
in the extended halos and then accreted back onto the galaxy.

To investigate this subject further, we present here new high resolution
calculations, addressing the ultimate fate of the ISM and SN ejecta,
and their mixing,
%The calculations we present here follow this line of research,
%addressing to the ultimate fate of the ISM and SN ejecta,
%and their mixing, 
in a realistic starbursting dwarf galaxy. We investigate in detail
the different phases of the gas flow, with particular emphasis
on the late evolution, evolving the simulations for 500 Myr
after the starburst event. We consider the effect of the dark matter,
gas rotation, thermal conduction and different starburst strengths. 
We also discuss the X-ray emission and its diagnostic for 
the abundance of the hot
gas, a particularly exciting topic in view of the forthcoming launch
of AXAF and XMM.

We aim at investigating the evolution of galactic winds in a general way,
without focusing on any specific object. Thus, we select the
parameters of the galactic models
(total mass, ISM mass and distribution, etc.)
to be representative of the class of dwarf galaxies.
Nevertheless, it can be useful to compare some of our results to
a real, representative object. An ideal galaxy is NGC 1569, a nearby,
well studied starburst galaxy.

Several independent lines of evidence indicate that NGC 1569 is
in a post-starburst phase (Israel 1988, Israel \& de Bruyn 1988, Waller 1991,
Heckman et al. 1995, Greggio et al. 1998),
with the major starburst activity ceased
$\sim 5 - 10$ Myr ago.
H$\alpha$ observations of NGC 1569 show (young) bubbles complexes,
filaments and arcs
throughout the volume of the galaxy (Tomita, Ohta \& Saito 1994),
suggesting a diffuse star formation.
Heckman et al. (1995) found that the H$\alpha$ emission
of NGC 1569 can be separated
in a quiescent component, permeating the starbursting region of the galaxy,
and a more violent component, far more extended and with velocities up
to 200 km s$^{-1}$. This high velocity component is interpreted to be
ionized shells of superbubbles and provides a direct evidence
of a galactic-scale outflow.

Heckman et al. (1995)
and Della Ceca et al. (1996) detected X-ray emission,
extending for 1-2 kpc along the optical minor axis of NGC 1569, thus
probing the hot gas phase directly.
This hot gas ($T\approx 10^7$ K) is the signature of the violent SN
activity on the ISM.

As in almost all studies to date, we make a number of simplifying 
assumptions in calculating our models. First,
the ISM is assumed to be homogeneous and single phase. Second, we neglect
the selfgravity of the gas, even if the gas mass is of the same
order of the stellar mass. Third, the starburst is instantaneous
and concentrated
in a small region at the center of the galaxy. While none of these
hypotheses is likely to be strictly correct, they allow for a more
direct comparison with previous works, and still make possible
the calculation of models retaining the basics attributes
of real galactic winds.
We will relax some of these assumptions in a future paper in preparation.

\section{Galaxy models}

Many ingredients play an important role in determining the
hydrodynamical evolution
of the galactic wind. Among others, the density distribution
of the ISM in the pre-burst galaxy, the energy injection rate
of the newly formed stars, the gravitational potential of the
galaxy and the effectiveness of transport processes in the gas,
like thermal conduction.

A thorough exploration of the parameter space would require an
enormous amount of computational resources and it is beyond the scope
of this paper.
Thus, we hold approximately constant the stellar and ISM masses
of the model galaxies ($M_*=1.7\times 10^8$ \msun and
$M_{\rm ISM} \sim 1.3 \times 10^8$ \msun), although $M_{\rm ISM}$ is
a crucial factor for the late evolution of the system
(De Young \& Heckman 1994; MF).
Instead, we vary some of the others parameters as
described below.

\subsection{The gravitational potential and the gas distribution}

The gravitational potential for our standard model is due
to two mass distributions:
a spherical quasi-isothermal dark matter halo plus a stellar
thin disk. 

The halo density is given by $\rho_{\rm h}(r)=\rho_{0h}/[1+(r/r_{\rm c})^2]$,
and we chose a central density $\rho_{0h} =
4.34 \times 10^{-25}$ g cm$^{-3}$ ($6.4 \times 10^{-3}$ \msun pc$^{-3}$).
The halo core radius is assumed to be $r_{\rm c}=1$ kpc.
The dark halo is truncated
at $r=20$ kpc. The total
dark matter mass is thus $\sim 2 \times 10^9$ \msun, while the halo mass
inside the galactic region (defined hereafter as a cylinder
$R<2.2$ kpc and $|z|<1.1$ kpc, approximately 
the optical size
of NGC 1569) is only
$0.66 \times 10^8$ \msun.

For simplicity, we assume that the stars are
distribuited in an infinitesimally thin Kuzmin's disk with
surface density
$$\Sigma_* (R)= {r_* M_*\over 2 \pi (R^2+r_*^2)^{3/2}}$$
where $r_*=2$ kpc is the radial scalelength and $M_* = 1.7 \times 10^8$
\msun is the total stellar mass, a typical value for dwarf galaxies.
Although this mass distribution is clearly a rough approximation
of real stellar disks, it does not degrade the accuracy of the
large scale hydrodynamical flow.
The stellar potential
generated by this mass distribution is
$$\Phi_*(R,z)=-{G M_* \over \sqrt{R^2+(r_*+|z|)^2}}$$
(Binney \& Tremaine 1987).

It turns out that the stellar mass inside the galactic region
is $M_{*,\rm gal} \sim 3.13 \times 10^7$ \msun, about half
of the dark halo mass and about a factor of four
less than the gas mass inside the same volume (see below).
The dark halo totally dominates the mass budget
at larger radii.

The ISM is assumed to be single-phase and in equilibrium
with the potential described above.
In real dwarf galaxies the neutral ISM is supported against
gravity partly by rotation
and partly by
the HI velocity dispersion (see Hoffman et al. 1996), with maximum rotational
velocity that typically exceeds the velocity dispersion by a factor of few.
Thus, in the standard model (hereafter model STD) 
we allow the ISM to rotate, to investigate the role
played by the angular momentum conservation on the late phase of
the evolution, when (once the energy output is ceased) the gas tends
to recollapse toward the central regions (see section 3.1).
The temperature of the unperturbed ISM is set to $T_0=4.5 \times 10^3$.
%corresponding to a line of sight velocity dispersion
%$\sigma = 5.3$ km s$^{-1}$.

To build a rotating ISM configuration in equilibrium with the given
potential, we first arbitrarily assume a gas distribution in
the equatorial plane ($z=0$) of the form
$\rho(R,0)=\rho_0/[1+(R/R_c)^2]^{3/2}$, where the central value
is $\rho_0=3.9 \times 10^{-24}$ g cm$^{-3}$
and the gas core radius is $R_c=0.8$ kpc.
The rotational velocity in the equatorial plane is then determined
from the condition of equilibrium:
$$ v_\phi ^2 = v_c^2 - {R\over \rho} \bigg | {dp \over dR} \bigg |_{z=0}$$
where $v_c=\sqrt{R d\Phi/dR}$ is the circular velocity and $p$
the thermal gas pressure. The rotational velocity is assumed to be
independent of $z$.
The density at any $z$ is then found integrating the $z$-component
of the hydrostatic equilibrium equation, for any $R$.
The edge-on and face-on profiles of the resulting gas column density are
shown in Fig. 1.
We note that this model, having an extended gaseous halo, 
resembles the models worked out by Silich \& Tenorio-Tagle (1998).

The circular velocity for this mass model increases with $R$,
reaching the maximum value of $\sim 20$ km s$^{-1}$ at $R\sim 4$ kpc
and staying almost constant for larger $R$. The rotational velocity
$v_\phi$ shows a similar radial behaviour, but with a maximum value
$v_\phi \sim 15$ km s$^{-1}$.

The total gas mass inside the galactic region is
$M_{\rm ISM,gal}\sim 1.32 \times 10^8$ \msun,
a typical
amount for dwarf galaxies (Hoffman et al. 1996), and
in close agreement with the mass inferred for NGC 1569 in particular
(Israel 1988). The total gas mass present in the numerical grid
(extending to 25 kpc in both $R$ and $z$ directions) is $M_{\rm ISM,tot}\sim
6 \times 10^8$ \msun.

The gas distribution qualitatively resembles that used by
Tomisaka \& Ikeuchi (1988). It has a low density region around 
the $z$-axis (see Fig. 2a),
which acts as a collimating funnel for the hot outflowing gas
(Tomisaka
\& Bregman 1993, Suchkov et al. 1994). This is due to the assumption
that $v_\phi$ does not depend on $z$. The funnel, however, 
influence the gas dynamics only at very large distances above the galactic
plane (i.e. for $z\gta 10$ kpc) and does not invalidate the results
presented in sections 3 and 4.

In order to address the influence of an intracluster medium (ICM) confining
the galactic ISM, we calculate model PEXT (section 4.3). In this simulation
we replace all the cold ISM (distributed as described above) having a thermal
pressure $P\le 10^{-13}$ dyn cm$^{-2}$ with a hot, rarefied ICM with
$\rho_{\rm ICM}=8 \times 10^{-30}$ g cm$^{-3}$ and $T_{\rm ICM}=10^8$ K.
In this case the cold ISM is confined to a roughly ellipsoidal region
with major and minor semiaxes 2 kpc and 1 kpc, respectively.
The galactic ISM mass is now only $M_{\rm ISM,gal}=1.05 \times 10^8$ \msun,
while the total mass of gas in the grid is $M_{\rm ISM,tot}\sim 1.16
\times 10^8$ \msun.

In addition to the models described
above, we use a different galaxy model (model B) to investigate the
effect of the absence of dark matter and rotation.  The isothermal ISM
in hydrostatic equilibrium in the
potential well generated by the same stellar distribution as in model
STD.
The central gas density is $\rho_0 =
1.1\times 10^{-23}$ g cm$^{-3}$, and the gas mass inside the galactic
region is $1.4 \times 10^8$ \msun, approximately as in model STD. 
Due to the lack of
rotational support, the gas distribution is now more concentrated than
in model STD (see the ISM column density in Fig. 1), 
and the total gas mass inside the grid is
$M_{\rm ISM,tot}=2.3 \times 10^8$ \msun.
We also run a model identical to model B, but including heat
conduction (model BCOND). 

\begin{figure}
 \vspace{6truecm}
 \caption{Column density of the initial ISM. Heavy solid line:
  model STD seen edge-on; heavy dashed line: model STD seen face-on;
  light solid line: model B seen edge-on; light dashed line: model B
  seen face-on.}
\end{figure}

\subsection{The starburst}

We assume an instantaneous burst of star formation which injects
energy in the ISM for a period of $30$ Myr, approximately the
lifetime of a $8$ \msun star, the smallest star producing a type II SN.
We consider two starburst strengths: the first is
representative of a moderate starburst, 
while the
second is intended to match more active galaxies.
We assume that the starburst produces a steady (mechanical) energy input rate
$L_{\rm inp}=3.76\times 10^{39}$ erg s$^{-1}$ (hereafter SB1 model) and
$L_{\rm inp}=3.76\times 10^{40}$ erg s$^{-1}$ (SB2 model) respectively.
%\footnote{Leitherer and Heckman consider all the energy output
%in calculating the energy deposition rate.
%However the real efficiency with which stellar wind and SN explosions
%heat the ISM is very uncertain, and can be much less than 1; see, e.g.
%Bradamante, Matteucci \& D'Ercole (1998).}. 
Model SB2 produces
a mechanical power similar to the lower limit estimated for NGC 1569
(Heckman et al. 1995).
The mass injection rate is assumed to be respectively
$\dot M=3 \times 10^{-3}$
\msun yr$^{-1}$ and $\dot M=3 \times 10^{-2}$ \msun yr$^{-1}$.

It is useful to compare the parameters we use with the detailed
starburst models of Leitherer \& Heckman (1995) (LH). For example,
for an instantaneous burst with a Salpeter IMF
(from 1 to 100 \msun) and metallicity $1/4$ of solar, they found
a mechanical energy deposition rate approximately constant
between 6 and 30 Myr after the starburst (see their fig. 55).
This justifies our
assumption of a steady energy source.
Our assumed mechanical luminosities for SB1 and SB2
correspond respectively to $\sim 2.1 \times 10^5$
and $\sim 2.1 \times 10^6$ \msun turned into stars during the
starburst event, according to LH.
If all stars with initial mass greater than 8 \msun end their lives
as type II supernovae, the total number of SNII events produced
by the starbursts is $\sim 4000$ and $\sim 40000$ for SB1 and SB2.

The total energy deposited after 30 Myr is $\sim 3.56 \times 10^{54}$
erg and $\sim 3.56 \times 10^{55}$ for SB1 and
SB2. These values must be compared with the binding energy of the gas
present in the numerical grid in the standard model,
$E_{\rm bind} \sim 5.3
\times 10^{54}$ erg, and with the binding energy of the gas inside the
galactic region $\sim 1.8 \times 10^{54}$ erg.

\begin{table}
  \caption{Models physical and numerical parameters.}
%%\begin{tabular}{@{}llrrrrrr@{}}
  \begin{tabular}{@{}llllllll@{}}
     & STD & SB1 & PEXT & B & BCOND \\
 $M_{\rm ISM}$  & 6.0  & 4.87  & 1.16  & 2.3  & 2.3  \\
 $M_{\rm ISM,gal}$ & 1.32  & 1.32  & 1.05  & 1.4  & 1.4  \\
 $M_{*}$  & 1.7  & 1.7  & 1.7  & 1.7  & 1.7   \\
 $M_{\rm DH}$ & 20  & 20  & 20  & 0  & 0  \\
 $L_{\rm inp}$ & $37.6$ & $3.76$ &
$37.6$  & $37.6$  & $37.6$ \\
 Rotation  & YES  & YES  & YES  & NO  & NO  \\
 $N_{\rm R}\times N_{\rm z}$ & $405^2$ & $505^2$ & $405^2$
& 480$\times$540 & 410$\times$530 \\
 $\Delta R_{\rm min}$ & $10$  & $3$  & $10$  & 
$2$  & $2$  \\
 $\Delta R_{\rm 2kpc}$ & $25$  & $17$  & 
$25$  & 22  & 22  \\
 $R_{\rm max}$ & 25  & 14.9  & 25  & 23  & 11.4      \\
 $z_{\rm max}$ & 25  & 14.9  & 25  & 42  & 38      \\
 Code          & ZEUS  & ZEUS  & ZEUS  & BO  & BO    \\

\end{tabular}
\medskip

$M_{\rm ISM}$ is the initial ISM mass present in the computational grid;
$M_{\rm ISM,gal}$ is the ISM mass in the galactic region 
$R<2.2$ kpc and $|z|<1.1$ kpc; $M_{*}$ is the assumed stellar mass;
$M_{\rm DH}$ is the dark matter halo mass. All masses are given in units
of $10^8$ \msun. $L_{\rm inp}$ is the energy input rate in $10^{39}$
erg s$^{-1}$. $N_{\rm R}$ and $N_{\rm z}$ are the number of cells in 
the $R$-direction and $z$-direction.
$\Delta R_{\rm min}$ (=$\Delta z_{\rm min}$) is the central
zone size in pc. $\Delta R_{\rm 2kpc}$ (=$\Delta z_{\rm 2kpc}$) is
the width in pc of the zone at (R,z)=(2 kpc,0) (or (R,z)=(0,2 kpc)). 
$R_{\rm max}$ and 
$z_{\rm max}$ are the total dimensions of the grid in kpc. In the last row
is indicated the hydrocode used (ZEUS-2D or the Bologna code).
 
\end{table}

After 30 Myr, the total mass returned to the ISM by
stellar winds and type II SNe
in the model by LH, again
with $Z=1/4$ $Z_\odot$  and a  Salpeter IMF, is $4.81 \times 10^4$ \msun
and $4.81 \times 10^5$ \msun for SB1 and SB2.
With our assumed $\dot M$, however, we inject $9 \times 10^4$ \msun
and $9 \times 10^5$ \msun for models SB1 and SB2 respectively.
Thus, we overestimate the mass return rate by a factor $\sim 2$
with respect to the LH model
\footnote{Alternatively, we can think to a starburst with a double
amount of mass turned into stars, and to an efficiency in the
energy deposition rate of $\sim 0.5$.}.
However, the hydrodynamical evolution of our models is not
sensitive to such a discrepancy, as well as our estimate of the
efficiency of ISM and metal ejection (although the pollution degree
of the ISM may be affected).

While our assumed starburst model is fairly consistent with
the detailed theoretical models by LH,
it is important to note that real galaxies have generally
a much more complex star formation history. The assumption
of instantaneous, point-like burst appears particularly severe.
For example, Greggio et al (1998) found that the bulk of the
starburst in NGC 1569 proceeded at an approximately constant star
formation rate of
$0.5$ \msun yr$^{-1}$ for $0.1 - 0.15$ Gyr (assuming a Salpeter IMF
from 0.1 to 120 \msun), until $\sim 5-10$ Myr ago,
when the star formation in the field ended.
It implies
that $\sim 5 - 7.5 \times 10^7$ \msun  of gas has been converted into
stars.

Moreover, H$\alpha$ observations of NGC 1569 show (young) bubbles complexes,
filaments and arcs distributed
throughtout the volume of the galaxy (Tomita et al. 1994),
suggesting a diffuse, wide scale star formation.

Galactic wind models powered by a point-like energy source are
nevertheless useful as first step toward the full complexity of the problem,
and for a direct comparison with previous studies.
Simulations with spatially and temporally extended star formation
will be the subject of a forthcoming paper.

\subsection{The numerical simulations}

To work out the models presented in this paper we used two
different 2-D hydrocodes. The first one has been developed
by the Numerical Group at Bologna Astronomical Observatory and
the (1-D) core of the scheme is described in Bedogni \& D'Ercole (1986).
This code and its successive extensions have been applied to a variety
of astrophysical problems (e.g. Brighenti \& D'Ercole 1997,
D'Ercole \& Ciotti 1998).
The second code employed is ZEUS-2D, a widely used, well tested scheme
developed by M. Norman and collaborators at LCSA (Stone \& Norman 1992).
We always found consistent results among the codes, as expected
from the numerous hydrodynamical tests performed with
the Bologna code (Brighenti 1992).

We solve the usual hydrodynamical equations,
with the addition of a mass source term and a thermal energy source term;
the hot gas injected expands to form the starburst wind with the appropriate
mechanical luminosity $L_{\rm inp}$.
%to represent the starburst wind.
These equations are described in details in, e.g., Brighenti \&
D'Ercole (1994).
The (constant) mass and energy source terms
are given respectively by
$\alpha =\dot M/{\cal V}$ and $\alpha \epsilon$. Here ${\cal V}$
is the volume of the source region, chosen to be
a sphere of radius=50 pc, centered at $(R,z)=(0,0)$,
and $\epsilon = L_{\rm inp}/ \dot M$.
To keep track of the gas lost by the stars formed in the starburst (the
ejecta), we passively advect it solving an ancillary continuity
equation for the ejecta density $\rho_{\rm ej}$.
Both the codes used spread shocks over 3-4 zones and contact
discontinuities over 4-10 zones.

In our models the angular momentum is treated in a fully consistent
way (see Stone \& Norman 1992 for the details about the
resolution of the angular momentum equation). Thus, contrary to
some of the previous studies (Tomisaka \& Ikeucki 1988, Tomisaka \& Bregman
1993), we do not use a reduced gravitational force
to mimic the rotational support of the ISM.

To take into account the thermal conduction (model BCOND)
we adopt the operator splitting method. We isolate the heat diffusion
term in the energy equation and solve the heat transport equation,
alternatively along the $z$ and $R$ direction separately, through the
Crank-Nicholson method which is unconditionally stable and second
order accurate. The system of implicit finite difference equations is
solved according to the two-stage recursion procedure (e.g. Richmeyer
and Morton 1967). Following Cowie \& McKee (1977), we adopt saturated
fluxes to avoid unphysical heat transport in presence of steep
temperature gradients.

We ran the models on a cylindrical grid (coordinates $R$ and $z$),
assuming axial symmetry. We use reflecting boundary conditions
along the axes and outflow boundary conditions at the grid edges.
To better resolve the central region,
the grid is unevenly spaced,
with the zone width increasing from the center to large radii.
Specifically, in the standard model (STD+SB2), the grid extends in both
the $R$-direction and $z$-direction from 0 to 25 kpc. The first zone
is $\Delta R = \Delta z = 10$ pc wide, 
and the size ratio between adjacent zones is 1.00747.

For the other models we use different grid spacing. For model SB1 (\S 4.1)
the inner grid size is 3 pc and the size ratio is 1.00717. 
For model B and BCOND the central zone is
only 2 pc wide and the size ratio between adiacent zones is 1.01.

The parameters used in the models and other characteristic quantities
are summarized in Table 1.

\section{The standard model (STD+SB2)}

\subsection{The dynamics of the ISM}

As the starburst wind starts blowing, the classical two shocks
configuration is achieved, in perfect analogy to the stellar wind
bubble theory (Dyson \& de Vries 1972, Weaver et al. 1977). The freely
expanding wind encounters the reverse shock and is heated to $T\sim
5 \times 10^7$ K, while the external shock sweeps the ISM. The shocked
starburst wind and the shocked ISM are separated by a contact discontinuity.
The reverse shock is always approximately spherical, since the short
sound crossing time in the shocked wind region keeps the pressure
almost uniform. The shape of the forward shock, instead, depends on the
ISM density distribution. The roughly oblate ISM configuration
of our model galaxy forces the superbubble to expand faster along the
polar direction, acquiring the classical bipolar shape at
late times (Fig. 2b,c,d). At earlier times, however, the density
distribution favours a diagonal expansion, generating
a curious boxy morphology (Fig. 2a).

When the energy input from the starburst ends (at $t=30$ Myr, Fig. 2a),
almost the whole galactic region
($R<2.2$ kpc, $|z|<1.1$ kpc)
is filled by the freely expanding wind,
a situation clearly unrealistic, due to our simple ISM model.
In real galaxies we expect
that this region hosts a complicated multiphase medium.
Israel \& van Driel (1990) found a relatively small hole (with
diameter $\sim 200$ pc) in the $HI$ distribution, associated
with the `super star cluster' N 1569A and likely caused by the
action of SNII and stellar winds of the star cluster.

The shocked ISM shell is accelerating through the steep
density gradient of the unperturbed ISM, and this acceleration promotes
Rayleigh-Taylor (R-T) instabilities. The shell tends to fragment
and relatively dense ($n\lta 0.5$ cm$^{-3}$), cold ($T\sim 10^4$ K)
filaments are clearly seen in Fig 2a
at $(z,R)\sim(1,3)$ kpc. The numerical resolution of this simulation
is not appropriate
to follow the real formation and evolution of these features,
whose actual density is expected to be higher than that found in our
computations.
As pointed out by MF, these
filaments of dense gas are immersed in the hot
gas and they do not necessarily trace the edges of the superbubble.

At t=30 Myr the whole shocked ISM shell is radiative.
Following the trend of the external ISM,
a density gradient is present along the shell, from the equator
to the pole, with the lateral side being denser ($n\sim 1.5$ cm$^{-3}$)
and the polar region more rarefied ($n\sim 0.002$ cm$^{-3}$).
At this time, the optical appearance
of the system,
if kept ionised (for example by a low level star formation activity,
hot evolved stars, etc.), would be that of an incomplete shell, the polar
portion being too rarefied to be observable, given its low emission measure
($EM\approx 0.01$ cm$^{-6}$ pc for the shell and $EM < 30$ cm$^{-6}$ pc
for the filaments).

The radiative external shock is too slow ($v_{\rm shock}\lta 180$ km s$^{-1}$)
to emit X-ray. This is contrary to the results by Suchkov et al. (1994),
who claim that most of the X-ray radiation comes from the shocked ISM.
The different behaviour in our models is due to the lower (by an order
of magnitude) $L_{\rm inp}$ considered, 
probably more appropriate for a dwarf galaxy. As explained in section 3.4,
we also find that the hot ISM is the most important contributor to the
X-ray luminosity. However, it is not heated by shocks, but by mixing
with the shocked wind at the contact discontinuities.
At later times the outer shock accelerates through the steep density
gradient, and heats the ISM to X-ray temperatures. On the other hand,
this happens only when the X-ray luminosity has dropped to very low
and uninteresting values (cf. Fig. 6).

\begin{figure}
 \vspace{12truecm}
 \caption{Map of the logarithm of the number density (cm$^{-3}$) 
  for model STD at four different times.
  The $z$-axis is vertical. Distances are in kpc. The gray-scale varies
  linearly from -7.65 (white) to 0.135 (black). Six contours are
  shown at levels -1, -2, -3, -4, -5 and -6.}
\end{figure}

Fig. 2b shows the density distribution at $t=60$ Myr. The steep
density gradient along the $z$-direction induces a radiative-adiabatic
transition of the polar portion of the external shock. The cold filaments
are slowly moving forward, and their density decreases to maintain
the pressure equilibrium with the expanding hot gas; now the densest
filaments have $n\sim 0.04$ cm$^{-3}$.

\begin{figure*}
 \vspace{11truecm}
 \caption{Density contours and velocity field for the central region
  of model STD at 140 Myr, 200 Myr, 250 Myr and 300 Myr.}  
\end{figure*}

The shell is increasingly thicker with time. In fact, while the outer edge is 
still expanding, with $v\sim 25$ km s$^{-1}$, the inner edge of the shell 
near the equatorial plane is already receding toward the center
with a velocity $v\sim 10$ km s$^{-1}$.
This backward motion, due to the drop of the pressure inside the
expanding hot bubble, will eventually cause the collapse of the cold
gas back inside the galactic region, as evident in Fig. 2d
(see also MF). The details of the collapse
are shown in Fig. 3 and described below.

Fig. 2c and 2d show the density at 100 and 200 Myr respectively.
The external shock assumes a prounounced cylindrical shape because
of the collimating effect of the low density region around the $z$-axis
(section 2.1). The polar
shock crossed the numerical grid edge (at $z=25$ kpc); however,
being the motion
supersonic, the numerical noise generated at the grid boundary does not
propagate back. 
Moreover, given the very low densities in that region,
the amount of gas lost from the grid is completely neglibile.

The temperature of the hot, X-ray emitting gas decreases with time.
$T\sim 5 \times 10^7$ K during the active energy injection phase
($t\lta 30$ Myr). At later times, radiative losses and especially
expansion lower the temperature ($T\lta 10^7$ K at 60 Myr; $T\lta 2
\times 10^6$ K at 100 Myr; $T\lta 10^6$ K at 200 Myr).  ASCA
observations of NGC 1569 (Della Ceca et al 1996) indicate that the
diffuse X-ray emission comes from a (luminosity weighted) $T\approx 8
\times 10^6$ K gas.  We note, however, that the temperature inferred
from simple fits to observed X-ray spectra may be a poor estimate of
the actual temperature (Strickland \& Stevens 1998). We warn that all
the model temperature values quoted are mass weighted and may not
represent the ``observable'' ones.

No more supported by the hot gas pressure,
the cold ISM recollapses at $t\sim 150$ Myr, filling again the
galactic region. We show a zoomed view of the central part of the grid
in Fig. 3. In panel a the density contours and the velocity
field are shown at $t=140$ Myr, just before the cold gas reaches
the center. In panel b the same quantities are shown at $t=200$ Myr.
In Fig. 3a the cold tongue is approaching the center at $v\sim 40$
km s$^{-1}$, with a Mach number ${\cal M} \approx 10$. The density
in the collapsing gas increases with $R$, from $n\sim 2.3 \times 10^{-3}$
cm$^{-3}$ at $(R,z)=(0.5,0)$ kpc, to $n\sim 7.5 \times 10^{-3}$ cm$^{-3}$
at $(R,z)=(1,0)$ kpc, to $n\sim 2.3 \times 10^{-2}$ cm$^{-3}$ at
$(R,z)=(2,0)$ kpc. At $t=150$ Myr the cold gas reaches the center
and shocks. Hereafter the accretion proceeds through a cylindrical
shock wave.

At $t=200$ Myr the cold gas is still flowing toward the center
with $v\sim 15$ km s$^{-1}$,
building a (transient) conical structure around the $z$-axis (Fig. 3b).
The mean density on the equatorial plane in the galactic region
is $n\sim 0.025$ cm$^{-3}$, and it is still growing with time.
Panels c and d show the subsequent evolution ($t=250$ Myr and $t=300$ Myr
respectively).
The collapse is the result of the
pressure drop in the hot bubble, due to its expansion along
the polar direction. Between $t=60$ Myr and $t=100$ Myr the
pressure of the hot gas decreases by more than one order of magnitude
(from $\sim 6 \times 10^{-13}$ to $\sim 10^{-15}$ dyn cm$^{-2}$).
The cold gas, no longer supported by the hot phase,
is driven back to the center mainly by its own pressure, rather than
the galactic gravity. 

We have verified that the collapse is not a spurious numerical
effect due to strong radiative losses at the numerically broadened
contact discontinuities. To this purpose we run a simulation identical
to the standard model, but with
the radiative cooling turned off.
For this adiabatic model we found that the collapse time
is again $\sim 150$ Myr.
The ISM recollapse
is thus an unavoidable phenomenon for all the models considered
in this paper.

%We show the time evolution of the pressure on the equatorial plane
%in Fig. 4 at two different times (60 Myr and 100 Myr). We also show the same
%quantities for the adiabatic models. 
%In both models the pressure in the hot phase
%decreases with time, although in the adiabatic model the dropping rate
%is slower. 

It is interesting to follow the circulation of the gas.
The mass of the ISM in the galactic region, at $t=200$ Myr, is
$M_{\rm ISM,gal}\sim 0.034 \times 10^8$ \msun,
about a factor of $\sim 40$ lower
than the initial
gas mass in the same volume (note that it does not mean that
the ejection efficiency $f_{\rm ISM}$ is 1/40,
since what matters in estimating
$f_{\rm ISM}$ is the amount of ISM effectively bound to the galaxy,
as described in the next section). However, at $t=200$ Myr the cold
gas is still flowing toward the center (Fig. 3b);
for instance, at $t=300$ Myr
we found $M_{\rm ISM,gal}=0.096 \times 10^8$ \msun.
At $t=500$ Myr,
the final time of our simulation,
$M_{\rm ISM,gal}=0.27 \times 10^8$ \msun, a factor of $\sim 5$ lower
than the initial gas mass in the same volume.

We can speculate further on the
fate of the cold gas falling back to the center at late times.
The face-on surface density of the central ISM
is an increasing function of time (Fig. 4),
since material continues to accrete until the end of our simulation
($t=500$ Myr; we could not follow the evolution further
because of our limitation in computational resources).
It has been suggested that above a critical ISM surface density
$\Sigma_{\rm crit}\sim 5-10 \times 10^{20}$
cm$^{-2}$, the star formation in dwarf galaxies is very efficient
(Gallagher \& Hunter 1984, Skillman 1987, 1996). At $t=200$ Myr
the face-on surface density peak in our model is
$\Sigma \sim 10^{20}$ cm$^{-2}$,
and it grows slowly to $6 \times 10^{20}$ cm$^{-2}$ at $t=500$ Myr.
Thus, we can hypothesize that the threshold surface density is reached
in a time of the order of 1 Gyr, after that a new burst of star formation
may start. This scenario is rougly consistent with
many studies of the star formation
history in BCD galaxies, which indicate that stars are formed
mainly through several discrete, short bursts, separated by long ($\sim$
few Gyrs) quiescent periods (see the review by Tosi
(1998) and references therein).

\subsection{The ISM ejection efficiency}

A key point in the galactic wind theory is the ability
of the starburst in ejecting the ISM (see Skillman \& Bender (1995)
and Skillman (1997) for a critical review about this subject).

\begin{figure}
 \vspace{6truecm}
 \caption{Face-on column density of model STD at three different times.
  Dotted line: 200 Myr; dashed line: 300 Myr; solid line: 500 Myr.}
\end{figure}

We estimate the ISM ejection efficiency calculating,
at some late time, for example t=200 Myr,
the mass of ISM $M_{\rm lost}$ which has velocity
{\it or} sound speed greater than the local escape velocity.
We assume that this gas (and the gas that already left the grid)
will be lost by the galactic system (see also MF).
It is important to note that $M_{\rm lost}$ calculated in this way
should be considered only a rough estimate of the amount of gas leaving
the galaxy, since dissipative effects may lower the ejection efficiency,
and the escape velocity depends critically on the poorly known size
of real dark matter halos.
The ISM ejection efficiency, $f_{\rm ISM}$, is then defined
as $M_{\rm lost}/M_{\rm initial}$, where $M_{\rm initial}$ is the
total gas mass present on the whole grid at $t=0$ (we neglect the contribution
of the ejecta, whose total mass is only $\lta 0.2$ \% of the
initial mass). We note that this operative definition for $f_{\rm ISM}$ is
grid dependent, since $M_{\rm initial}$ increases with the volume
covered by the numerical grid.

At t=200 Myr we find
$f_{\rm ISM}=0.058$: evidently even a powerful starburst
as the one considered for this model is not effective in removing
the interstellar gas.

However, as pointed out in the previous section, the gas mass
inside the galactic region
can be significantly lower than the initial value, even long after the
starburst event: thus, {\it the efficiency in removing the
ISM from the central regions may be considerably greater
than $f_{\rm ISM}$.} However, for other models (see section 4.3),
the galaxy is able to recover most of the original ISM mass
after $\approx 100$ Myr.

\subsection{The enrichment}

In a similar way we have estimated the
ejection efficiency of the metal-rich stellar ejecta,
$f_{\rm ej}$.
At $t=200$ Myr we found $f_{\rm ej} = 0.46$: {\it the galaxy
is less able to retain the enriched stellar ejecta than its
own original ISM}. This finding supports the selective winds
hypothesis, and it is in
qualitative agreement with others numerical simulations
(De Young \& Gallagher 1990; De Young \& Heckman 1994;
MF).

It is interesting to investigate the spatial distribution of the
ejecta material. We found that at $t=200$ Myr $\sim 7.3 \times 10^5$
\msun of stellar ejecta are present on the numerical grid,
about 80 \% of the total material released by the starburst.
However, the ejecta mass in the galactic region is only
$M_{\rm ej,gal}\sim 5.15 \times 10^3$ \msun, less than 0.6 \% of the total
amount ejected ($9 \times 10^5$ \msun)!
Since gas continues to flow toward the central region, the mass
of the ejecta in the galactic region increases slightly with time.
At $t=300$ Myr, for instance, we found $M_{\rm ej,gal}\sim 1.29 \times 10^4$
\msun, and $M_{\rm ej,gal}\sim 3.6 \times 10^4$ \msun at $t=500$ Myr.
We conclude that, {\it while a significant fraction of the
ejecta is retained by the relatively deep potential of the dark halo,
most of it resides in the outer regions of the system, in a
phase so rarefied to be virtually unobservable}.

The cold gas collapsing at late times, and filling the galactic region,
has been only slightly polluted by the stellar ejecta. To characterize
the pollution degree we introduce the local ejecta fraction as
${\cal Z} = \rho_{\rm ej}/\rho$, where $\rho_{\rm ej}$ is the density
of the ejecta. The average ejecta
fraction in the galactic region, at $t=200$ Myr, defined as
$ <{\cal Z}_{\rm gal}>=M_{\rm ej,gal}/M_{\rm ISM,gal}$, is 
${\cal Z} \sim 1.4 \times 10^{-3}$.
The cold galactic ISM, probably the only component detectable at late times
because of its relatively high density,
shows only a small degree of enrichment.

We can estimate the increase in the metal abundance generated by the
starburst from the total number of SNII, which we assume to be the only
source of metals.
%(i.e. we neglect the contribute by SNIa and
%stellar winds). This assumption is justifed for
%$\alpha$-elements (like oxygen), synthetized mainly inside
%massive stars. Nevertheless, an investigation
The iron production and circulation is particularly worthwhile,
because the metallicity estimated through X-ray spectra of
the hot gas phase ($T\sim 10^7$ K)
are especially sensitive to iron through the Fe-L complex at $\sim 1$ keV.
For the sake of simplicity we shall neglect the iron produced by SNIa,
whose iron release timescale is believed to be
of the order of one Gyr (Matteucci \& Greggio 1986), a time
much longer than those considered in this paper.

In section 2.2 we estimated a total number of SNII $\sim 4000$ and $\sim 40000$
for SB1 and SB2 respectively (adopting the same IMF as in LH). 
%These numbers depend on the IMF assumed, and
%we followed LH, choosing a Salpeter IMF
%with lower and upper masses of 1 and 100 \msun.
The yields of metals from SNII are rather uncertain, especially
for iron and oxygen, because of the complications in the late evolution
of massive stars and nuclear reactions rates.
A compilation of IMF averaged yields, i.e.,
the mean ejected mass of a given element per SN,
can be found in Loewenstein and Mushotzky (1996).
Given the approximate nature of the calculations presented in this paper,
we simply adopt $<y_{\rm Fe}>=0.1$ \msun  and $<y_{\rm O}>= 1$ \msun
as reasonable values for averaged iron and oxygen yields.
%The total iron and oxygen masses produced by the starburst are thus
%$M_{\rm Fe}=4 \times 10^3$ \msun and $M_{\rm O}=4 \times 10^4$ for SB2.
We assume that the metals are well mixed within the ejecta, whose
abundances (by mass and relative to H) 
are $Z_{\rm Fe,ej}\sim 3.4$ $Z_{\rm Fe,\odot
}$ and
$Z_{\rm O,ej}\sim 4.6$ $Z_{\rm O,\odot}$, where we adopt
the meteoritic solar
abundances from Anders \& Grevesse (1989).

In Fig. 5 we show the iron gas abundance distribution at $t=200$ Myr, assuming
that the original ISM has $Z_{\rm Fe}=0$ (i.e. we calculate
the increment in the metallicity caused by the starburst ejecta).
The iron abundance is highly inhomogeneous, both in the hot and cold
phase. It ranges from very low values $Z_{\rm Fe}\lta 0.01$
$Z_{\rm Fe,\odot}$, to the pure ejecta value $Z_{\rm Fe}=3.4$
$Z_{\rm Fe,\odot}$. The hot phase metallicity is supersolar with
typical values $Z_{\rm Fe}=1.5-2.5 \;\; Z_{\rm Fe,\odot}$.
It is puzzling that recent ASCA observations of the outflows
in starburst galaxies indicate that the metal abundance 
of the hot gas is rather low. We discuss this point in the next section.

\begin{figure}
 \vspace{8.5truecm}
 \caption{Distribution of the iron abundance, calculated as explained
  in the text, for model STD at $t=200$ Myr. 
  Labels are in kpc. Minimum and maximum values, in solar units, are 0 (white)
  and 3.4 (black). Contours are shown at $Z_{\rm Fe}=0.5,1,1.5,2,2.5,3$.}
\end{figure}

While the numerical diffusion may affect somewhat
the absolute values of $Z_{\rm Fe}$,
we believe that the spatial variations of the metallicity are real.
The oxygen abundance pattern is identical to that of iron,
due to our assumption of perfectly mixed ejecta, but with different
minimum and maximum values ($0-4.6$ $Z_{\rm O,\odot}$).

The cold gas replenishing the galactic region has average metallicity
$<Z_{\rm Fe}>\lta 0.005$ $Z_{\rm Fe,\odot}$ 
($<Z_{\rm O}>\sim 0.01$ $Z_{\rm O,\odot}$),
so that a successive instantaneous starburst
event would form stars only slightly more metallic than the previous stellar
generation.

\subsection{The X-ray emission and hot gas metallicity}

A detailed investigation of the X-ray emission of the hot gas is
beyond the scope of this paper. The intrinsic diffusion of the
numerical scheme spreads contact discontinuities, separating the hot
and cold phases, over several grid points. This fact prevents us to
consistently calculate X-ray luminosities ($L_X$) and emission
averaged abundances of the hot phase. In fact, the gas inside the
broadened contact discontinuities (a mixture of the ejecta and the
pristine ISM), being
relatively dense and with temperatures of the order of $10^6$ K,
turns out to dominate $L_X$.

It is important to note that several physical processes, not considered
in this simulation, smear out hydrodynamical discontinuities,
mixing cold and hot ISM, and producing a gas phase
with intermediate temperature
and density. Most important are thermal conduction
and turbulent mixing (Begelman \& Fabian 1990).
Thus, the undesired numerical diffusion qualitatively mimics real
physical effects. We consider explicitely the heat conduction
in section 4.3, but we are not in the
position to make quantitative estimates
on the influence of turbulent mixing layers on $L_X$.

With this limitation in mind we can nevertheless gain some insights on
the properties of the X-ray emission of starbursting galaxies. The
X-ray luminosity of several models (see below), 
calculated in the straightforward way as
$L_X=\int{\epsilon_{R}(T) dV}$, is a decreasing function of time
(here $\epsilon_{R}(T)$ is the Raymond-Smith emissivity in the
{\it ROSAT} band). All models shown in Fig. 6 share the same trend: at first
$L_X$ drops gently until $t=30$ Myr and then, when the energy input stops,
$L_X$ decreases rapidly to unobservable values.

\begin{figure}
 \vspace{7.5truecm}
 \caption{X-ray luminosity vs. time for four models. Solid line:
  model STD; dashed line: model SB1; dotted line: model B;
  dot-dashed line: model BCOND.} 
\end{figure}

One of the most interesting observables is the emission averaged
metallicity. In order to compare our data with oservations, we define
the emission averaged ejecta fraction
$<{\cal Z}>_X=(1/L_X) \int{{\cal Z}\,\epsilon_{R}\,dV}$
(it is a measure of the gas
abundance \footnote{The emission averaged iron abundance is $Z_{\rm Fe,ej}
< {\cal Z}>_X$, where $Z_{\rm Fe,ej}=3.4$ $Z_{\rm Fe,\odot}$ with the
assumptions adopted in section 3.3.}). 
It is almost constant with time up to 30 Myr, 
with typical values of 0.04-0.06,
, and then it increases steadly up to
$\sim 0.5$ at $t=100$ Myr, when, however the X-ray luminosity is
so weak to make the detection virtually impossible. 
This very low `metallicity' clearly indicates that most of
the X-ray emission comes from original ISM mixed with stellar ejecta.

To isolate the contribution of the ejecta material to $L_X$ and 
$<{\cal Z}>_X$, we recalculated these two quantities using the ejecta
density $\rho_{\rm ej}$ (instead of the gas density) in the calculation 
of the X-ray emissivity. Now $L_X$ is a factor $\sim 100$ lower
than before, and $<{\cal Z}>_X$ is now much larger, approaching unity.
Note that this does not mean that the ISM heated by the outer shock
is the main contributor
to $L_X$ (see below). It demonstrates instead that diffusion processes
mix the ISM with the shocked ejecta, and the material in these mixing
layers makes most of the $L_X$.

In the last few years the {\it ROSAT}, {\it ASCA} and {\it BeppoSAX}
satellites provided
detailed X-ray observations of starburst galaxies. While for dwarf galaxies
the hot gas abundance cannot be unambiguosly determined (e.g. Della Ceca
et al. 1996), for brighter starburst galaxies the counts statistics is
high enough to make this task possible (Ptak et al 1997; Tsuru et al.
1997; Okada, Mitsuda \& Dotani 1997; Persic et al. 1998).
A somewhat surprising result of all these observations is that the iron
abundance is invariably small, typically less than 0.1 solar.
This low metallicity can easily be understood if the X-ray emission
is dominated by the layer of shock-heated ISM, 
as pointed out by Suchkov et al. (1994).
However, this is not a general result, and it does not hold for our models
in particular, since the external shock
is too slow to heat the ISM to X-ray temperatures (section 3.1).
Thus, in model STD the only X-ray emitting gas is expected to be the
(shocked) ejecta of the stars formed in the starburst, and its metallicity is
thus expected to be quite high.
This abundance discrepancy forces
the theoretical models to move toward a higher level of complexity.

Low X-ray abundances can be explained in several ways.
First, it seems reasonable that thermal conduction and turbulent mixing
give rise to a mass loaded flow (Hartquist, Dyson and Williams 1997,
Suchkov et al. 1996) with low emission averaged metallicity, provided
that the cold gas mixed with the hot phase is nearly primordial. In
this case the emission averaged temperature of the hot gas is expected
to be low (few $10^6$ K); see section 4.3.
Second, the hot gas might be severly depleted by dust. Stellar outflows
and SN ejecta
are observed to form dust (e.g. Clegg 1989; Colgan et al. 1994), and the
dust sputtering time $t_{\rm sp} \sim 2 \times 10^6 a_{\mu {\rm m}}/n $ yr
( where $a_{\mu {\rm m}}$ is the dust grain radius,
Draine \& Salpeter 1979; Itoh 1989) in
the hot phase
may be long enough to make most of the iron still locked into grains
after few $10^7$ yr.
Another possibility is that the estimated abundances are not accurate.
Strickland \& Stevens (1998) analysed the synthetic {\it ROSAT}
X-ray spectrum of a
simulated wind-blown bubble, finding that simple fits
may underestimate the metallicity by more than one order of magnitude.
The inadequacy of 1-T models in estimating the gas abundance has been
demonstrated also by Buote \& Fabian (1998) and Buote (1999) in the
context of hot gas in elliptical galaxies and groups of galaxies.
Indeed, Dahlem, Weaver \& Heckman (1998) used multi-components 
models to fit {\it ROSAT}
PSPC + {\it ASCA} spectra of seven starburst galaxies and found that
low metallicities are no more required, and
nearly solar abundances are entirely consistent with the data.
Their findings support the idea that the inferred low abundances
are caused by the undermodelling of X-ray spectra.

\section{Other Models}

\subsection{Model SB1}

With this simulation we investigate the effect of a weaker starburst
on the ISM of a dwarf galaxy.
This model is identical to model STD, but the starburst
mechanical luminosity is a factor of ten lower. This starburst may be more
typical among dwarf galaxies. We reduce the mechanical luminosity
lowering the mass loss rate by a factor of ten (see \S 2.2).

%For this simulation we used a finer numerical grid (with respect to
%model STD). Since the power injected by the starburst is lower than
%in model STD, the
%superbubble will be correspondingly smaller, and higher resolution is
%needed to appropriately resolve the interesting features.
%As before, the grid is non-uniformly spaced, with the zone size
%increasing from the center to the outer boundary.
%The smallest zone is now 3 pc wide, and the grid extends
%to $z_{\rm max}=R_{\rm max}=15$ kpc and contains 505 $\times$ 505 zones.

We anticipate that in this model the radiative cooling at the contact
discontinuities, artificially broadened by numerical diffusion, is
now important, and causes the hot bubble to slowly collapse.

Fig 7a shows the density distribution at $t=30$ Myr, when the energy and
mass input turns off. As in model STD, dense
tongues of shocked ISM penetrate in the hot bubble
as a result of R-T instabilities.
As expected,
the superbubble is now much smaller than in model STD, and it is
expanding less rapidly. The hydrodynamical evolution is illustrated
in the other panels of Fig. 7. At $t=60$ Myr (Fig. 7b) the internal edge
of the shocked ISM shell is receding toward the center
with $v\sim 25$ km s$^{-1}$. The cold gas reaches the origin at $t\sim 70$ Myr,
much earlier than in model STD.
We find that the ISM collapse is slightly
accelerated by the spurious energy losses mentioned above. To address
the importance of this undesired numerical effect, we recalculated the
Model SB1 without radiative cooling (the adiabatic model). 
These two extreme models
should bracket the reality. For this adiabatic model
the replenishing of the central region occurs at $t\sim 85$ Myr.

Fig. 7c shows the ISM density at $t=100$ Myr. The hot, rarefied bubble
is almost totally shrunk; the hot gas mass is now only $4.4 \times 10^2$
\msun (it was $\sim 2.7 \times 10^4$ \msun at $t=30$ Myr). At the same time
the adiabatic model contains $\sim 2.6 \times 10^4$ \msun of hot gas.
The cold ISM continues to move ordinately
toward the $z$-axis, and it encounters
a weak accretion shock at $R \sim 0.1$ kpc.

\begin{figure}
 \vspace{12truecm}
 \caption{Map of the logarithm of the number density 
  for model SB1 at four different times.
  Labels are in kpc. The gray-scale varies
  linearly from -5.42 (white) to 0.30 (black). Contours are as in Fig. 2}
\end{figure}

The density
distribution at 200 Myr is shown in Fig. 7d. No more hot gas is
present (while in the adiabatic model $M_{\rm hot}\sim 4.8 \times 10^3$
\msun, and it is decreasing
with time as the result of the numerical diffusion).
The accretion shock has moved forward to $R \sim 1$ kpc, where the cold ISM
is still accreting with $v\sim 10$ km s$^{-1}$.
The face-on surface density in the galactic region varies from $2.5 \times
10^{21}$ cm$^{-2}$ at the very center, to $4 \times 10^{20}$
cm$^{-2}$ at $R=2$ kpc. The number density in the central region
is about $0.35$ cm$^{-3}$. As in model STD the central surface density
is slowly increasing with time, approaching the critical value for
the onset of effective star formation activity. Thus, also for this model,
the secular hydrodynamical evolution indicates the possibility of
recurrent starburst episodes.
The time between successive starburst events in this model
is shorter that in model STD, being only few 100 Myr.

At $t=200$ Myr the ISM ejection efficiency
$f_{\rm ISM}$ is essentially zero:
all the gas is cold and it is moving
with a velocity lower than the escape velocity ($f_{\rm ISM}=1.8 \times 
10^{-3}$ for the adiabatic model).
The gas mass inside the galactic region is $M_{\rm ISM,gal}=6.5 \times 10^7$
\msun, about half of the mass present initially. Since the gas is still
accreting, the central ISM mass increases with time: at $t=300$ Myr
we have $M_{\rm ISM,gal}=8.0 \times 10^7$ \msun.

Thus, in the case of moderate
starburst strength, the galaxy is able to recover most of the
original ISM in a relatively short time.
The evolution of this model is qualitatively similar to that of model STD,
but is now accelerated and, as expected, the galactic ISM `forgets'
the starburst quicker.

The circulation of the stellar ejecta is qualitatively similar
to that of the standard model. However, now $f_{\rm ej}=0.003$:
almost all the metals produced by the starburst remain bound to the galaxy.
A significant fraction of the total ejecta mass ($\sim 2.4 \times 10^4$
\msun, $\sim 27$ \% of the total)
is still present in the galactic region at this late time.
The very low value for $f_{\rm ej}$ is partly due to the excess of 
radiative losses at the contact surfaces. For the adiabatic model
we find $f_{\rm ej}=0.14$ (still much lower than in model STD)
and $M_{\rm ej,gal}\sim 3.3 \times 10^4$ \msun. In summary, we find that
$f_{\rm ej}$ is significantly lowered by the spurious extra-cooling,
but the important quantity $M_{\rm ej,gal}$ does not change greatly.
The conclusion is that a significant fraction
($\approx 30$ \%) of the metals ejected is retained in the
galactic region
when the moderate starburst SB1 is adopted.

\begin{figure*}
 \vspace{8truecm}
 \caption{Map of the logarithm of the number density
  for model PEXT at three different times.
  Labels are in kpc. The gray-scale varies
  linearly from -7.27 (white) to 0.20 (black). Contours are as in Fig. 2}
\end{figure*}

\subsection{Model PEXT+SB2}

With this model we investigate the evolution of a galactic wind
occurring in a galaxy immersed in a hot, tenuous ICM as described in
section 2.1. All the other parameters are identical to model STD.
Fig. 8a shows the gas density at $t=30$ Myr. The superbubble has
already blown out in the ICM, generating a complex filamentary
structure. The fastest material penetrating in the ICM is moving
with $v\approx 2000$ km s$^{-1}$. Fig. 8b and 8c show the density
at $t=60$ Myr and at $t=200$ Myr. The portion of the
cold shell blowing out in the ICM 
is completely disrupted by the instabilities
and spreads in a large volume, due to the high expansion velocities
in the rarefied medium.
At 200 Myr, the original ISM survives in a toroidal structure
($1.5 < R < 8$ kpc) on the equatorial plane. 
The inner edge of the
cold gas is receding slowly ($v\sim 20$ km s$^{-1}$) 
toward the center, while the outer portion
is still expanding ($v\sim 40$ km s$^{-1}$).
The cold gas
starts to collapse toward the center, which is reached at $t\sim 270$ Myr,
much later than in the previous models.
The ISM column density increases more slowly than in model STD,
and at $t=500$ Myr the central peak is only $\Sigma_0\sim 2
\times 10^{20}$ cm$^{-2}$.
Thus, in this case, the subsequent star formation episode might be delayed
with respect to model STD. 

At 200 Myr the mass of gas present in the galactic region
is $M_{\rm ISM,gal}\sim 1.9 \times 10^6$ \msun, and about 1.5 \% is
hot ($T\sim 10^6$ K). At the final time (500 Myr) we have 
$M_{\rm ISM,gal}\sim 8.7 \times 10^6$ \msun.
The ejection efficiency is $f_{\rm ISM}=0.31$, much higher than in model STD
because the absence of an extended envelope of (relatively dense) cold
gas.

The mass of the metal-rich ejecta in the galaxy
is $M_{\rm ej,gal}\sim
5.8 \times 10^3$ \msun ($1.8\times 10^4$ \msun at 500 Myr)
and $f_{\rm ej}=0.83$.
These values are comparable to those found for model STD. However,
we find that the hot gas has been severly contamined by the hot ICM,
and ${\cal Z} \lta 0.05$ for almost all the hot ISM.
The reason for this behaviour is the high temperature of the ICM,
which greatly increases the importance of numerical diffusion.

We estimate an upper limit for this effect, considering the first order
upwind method (Roache 1972). The numerical diffusion coefficient is
$D_{\rm upwind}\approx c \Delta$, where $\Delta$ is the zone size.
The diffusion time is $\tau_D=\Delta^2/D \approx 30$ Myr
(here $\Delta\approx 30$ pc at $R=z\sim 2.5$ kpc and $c$
is the ICM sound speed),
so the numerical diffusion affects significantly this simulation, and
this explains the very low values for ${\cal Z}$. We conclude that
for model PEXT we cannot calculate the enrichment process in a
consistent way. For model STD, given
the low temperature of the ISM ($4.5 \times 10^3$ K),
$\tau_D$ is more than two order of magnitude longer, and the intrinsic
diffusion is negligible.

We note that the {\it physical} diffusion time scale, $\tau_D=L^2/D$, where
$L$ is the typical length scale of the problem ($L\approx 1$ kpc), 
is very short:
$\tau_D\approx 10^2 - 10^3$ yr. This is due to the high value for $D\approx
\lambda c$, where $\lambda\approx 5$ Mpc is the mean free path
for the ICM (Spitzer 1962). However, even a small
magnetic field reduces the mean free path to the
order of the ion Larmor radius $r_L$. Only in this case we are allowed
to consistently use the hydrodynamical equations. With $\lambda\approx r_L$
the physical diffusion is effectively impeded.

\subsection{Model B+SB2 and BCOND+SB2}

Panel a of Fig. 9 shows the gas density of model B (section 2) 
at 30 Myr, just at the
end of the starburst activity.
The free expanding wind extends so far that almost
all of the galaxy is devoid of the pristine gas. There is a radial gradient
in the bubble temperature: along the $z$-axis $T$ ranges 
from $\sim 4\times 10^7$ K
close to the reverse shock to $\sim 10^6$ K behind the forward shock;
a similar pattern is also present along the
$R$-axis, although the temperatures behind the lateral
shock are lower ($\sim 10^5$ K) because of the lower velocity of the shock
moving through the higher local ambient density. The average density
of the hot gas filling the bubble 
is $\approx 10^{-4}$ cm$^{-3}$. The expansion
velocity along the symmetry axis is $\sim 300$ km s$^{-1}$, and decreases
toward the equatorial plane. The bubble accelerates
as it expands through the decreasing ISM density profile, and
the R-T unstable contact discontinuity generates relatively dense ($n\sim
10^{-2}$ cm$^{-3}$) and cold ($T\sim 10^4$ K) filaments and
blobs. Actually, denser structures can be seen on the $z$ axis, but they
are likely due to our assumption of vanishing radial velocity on the
symmetry axis. In fact, cold gas deposited on this axis cannot be
effectively removed, a well known shortcoming common to all 2D
cylindrical simulations. Given the progressively
increasing zones size with $z$ and $R$, it is likely that the knots
density is underestimated in our simulations, especially for the
condensations far from the center.
At $t=57$ Myr (Fig. 9b) there is a large region essentially
devoid of gas ($n\sim 5\times 10^{-6}$ cm$^{-3}$) surrounded by a very
thick, low density shell ($n\sim 10^{-4}$ cm$^{-3}$), with a
temperature $\sim 10^6$ K. The external shock is rounder than in
model STD, due to the lack of the collimating effect of the funnel
along the $z$-axis (cfr. Fig. 2b).
The dense and cold gas near the equatorial plane is already receding toward the
center with a velocity $\sim 30$ km s$^{-1}$,
and a rarefaction wave is moving outward. The highest density in the shell
is $\sim 10$ cm$^{-3}$ on the equator, where the
expansion velocity is $10$ km s$^{-1}$. Apart the cold gas on
the $z$ axis, where the density reaches $\sim 30$ cm$^{-3}$, the
densest filaments have $n\sim 3 \times 10^{-2}$ cm$^{-3}$. At $t\sim 75$ Myr
(not shown in Fig. 9) the inflowing cold gas reaches the center, 
where the density
is still rather low ($n\sim 10^{-2}$ cm$^{-3}$).
s$^{-1}$. After 106 Myr (panel c), the final time of this simulation,
the hot gas is still expanding, but
the central cold ISM is enterely collapsed, filling a region
$|z|<2.5$ kpc, $R<6.5$ kpc. The galaxy has thus recovered a cold ISM 
distribution similar
to the original one. The ISM mass inside the galaxy is $M_{\rm ISM,gal}
\sim 10^8$ \msun.
Near the center the density reaches a few cm$^{-3}$.

\begin{figure*}
 \vspace{8truecm}
 \caption{Map of the logarithm of the number density 
  for model B at three different times.
  Labels are in kpc. The gray-scale varies
  linearly from -6.31 (white) to 0.76 (black).}
\end{figure*}

\begin{figure*}
 \vspace{8truecm}
 \caption{Map of the logarithm of the number density 
  for model BCOND at three different times.
  Labels are in kpc. The gray-scale varies
  linearly from -6.33 (white) to 0.95 (black).}
\end{figure*}

About the starburst ejecta, its largest content inside the galaxy is reached
at $t=30$ Myr, with $M_{\rm ej,gal}=3.6\times
10^5$ \msun. At this time $<{\cal Z_{\rm gal}}>=2.4\times
10^{-3}$, and it decreases steadly with time. At $t=106$ Myr
$<{\cal Z_{\rm gal}}>=1.1\times 10^{-3}$ while the ejecta mass is
$M_{\rm ej,gal}\sim 10^5$ \msun, about one tenth of the total gas lost by
the massive stars. 
The ISM and metals ejection efficiency, estimated at $t=106$ Myr, are
$f_{\rm ISM}=0.48$ and $f_{\rm ej}=0.77$ respectively.
While the high $f_{\rm ej}$ is consistent with the results obtained
in section 3.3,
the large value for $f_{\rm ISM}$ is striking when compared to model STD.
However, this discrepancy reflects a deficiency in the definition
of $f_{\rm ISM}$, rather than a really different behaviour of the two
models.
As a matter of fact, model B recovers a `normal' ISM {\it before} model STD!
The discrepancy is due to several factors. First, in model B the massive dark
halo is absent. The escape velocity is then quite low (a factor 2-4 less
than in model STD) and the gas residing at large radii becomes easily
unbound. Second, the ISM distribution in model B is more peaked than
in model STD, and the total amount of gas present in the numerical
grid is lower (cfr. section 2.1). This, in turn, means that in model B
the starburst provides more energy per unit gas mass than in model STD.
We believe that the difference in $f_{\rm ISM}$ between model STD
and model B should be considered with some caution.
In real galaxies, the gas at large radii, which is the source of
the difference in $f_{\rm ISM}$, can be removed by 
ram pressure and tidal stripping,
processes not included in our simple models. Thus, the contribution
of this gas to $f_{\rm ISM}$ is rather uncertain.
%In real galaxies, the fate of the gas at large radii, which is the source of
%the difference in $f_{\rm ISM}$, likely depends on its interaction
%with the environment, being subject to ram pressure and tidal stripping,
%processes not included in our simple models.

The X-ray emission averaged $<{\cal
Z}>_X$ is much higher than $<{\cal Z_{\rm gal}}>$ and $increases$ from $<{\cal
Z}>_X=0.06$ at $t=30$ Myr up to $<{\cal Z}>_X\sim 0.2$ at $t\gta 50$ Myr; at
later times the bubble gas cools out of the X-ray temperatures and $<{\cal
Z}>_X$ drops to zero at $t\sim 60$ Myr. 
%The reason
%of the initial increase of $<{\cal Z}>_X$ at the end of the starburst
%is due to the recollapse of the cold gas; a larger portion of it is now
%inside the galaxy. As pointed out in section 3.4, a large fraction of
%the X-ray luminosity comes from the edges of the hot bubble which thus
%are crucial of $<{\cal Z}>_X$. The hot gas outside the galaxy has a
%much higher ${\cal Z}$, but it has an expansion velocity higher than
%the escape velocity and is definetively lost.

In Fig. 10 we show model BCOND, identical to model B
but with the heat conduction activated. Again, panel a
shows the density at 30 Myr. The
superbubble is less extended than in model B, because of the increased
radiative losses in the conduction fronts.
The temperature distribution inside the superbubble 
is now rather flat near the equator,
but a negative gradient is present along the $z$-direction, with the
temperature in the range $10^6<T<10^7$ K. As in the previous
models, cold structures
are present due to the R-T instabilities. The density
of these structures is $n\sim 3$ cm$^{-3}$, while the density of the
hot gas is $\sim 10^{-3}$ cm$^{-3}$, one order of magnitude larger than in
model B. This higher density is
due to the evaporation of the walls of the shocked ISM shell 
which `feed' the inner region
of the bubble. The expansion velocities are
similar but lower than those of model B at the same
time. Panel b shows the density at 56 Myr and can be compared
with panel b of Fig. 9. The size of the superbubble remains
smaller and the shape more elongated.
The hot gas in the cavity is denser ($n\sim
5\times 10^{-5}$ cm$^{-3}$) and slightly colder ($T\sim 2.5\times 10^5$ K)
than in the non-conductive case.
The cold gas near the equatorial plane
is receding toward the center, while the outer edge (where $n\sim 10$ 
cm$^{-3}$) is still expanding. The densest filaments have
$n\sim 0.1$ cm$^{-3}$. Panel c shows the gas flow at 105 Myr.
The cold inflowing gas has just reached the center,
much later than model B.
In fact, the pressure drop of the hot gas
is slower than in model B because the density of the hot gas
is kept higher by the shell evaporation and by 
the slower expansion rate. 

At $t=125$ Myr, the last time of this simulation, $M_{\rm ISM,gal}\sim
0.92 \times 10^8$ \msun, not far from the initial value. However, 
only a fraction of the galactic volume contains a cold, dense ISM.
In fact, roughly half of the galaxy is still filled with the rarefied gas
of the cavity, now only moderately hot ($T\lta 10^5$ K).
At this time the ISM ejection efficiency is $f_{\rm ISM}=0.32$.

The peculiar structure apparent on the simmetry axis ($z\gta 20$ kpc) at
$t=105$ Myr (Fig. 10c) is a numerical artifact 
depending on our treatment of the
heat conduction. Collisionless shocks (for istance, in supernova
remnants) do not show the hot precursor which would be expected
(Zel'dovic \& Raizer, 1966). This means that the plasma
instabilities responsible of the shock front formation also inhibit
the heat flow through the front itself
(Cowie 1977). To mimic this phenomenon in
numerical simulations, the heat conduction coefficient must vanish at
the shock front. To detect the shock front position on the computational grid
is an easy task in 1D simulations, but becomes rather cumbersome in
two dimensions. Fortunately, the precursor length is rather short
(shorter than the grid size) unless the upwind density is very low.
We thus did not make any special treatment at the shock front.
Effectively, the heat flux overruns the front only at late times, when
the upwind density becomes rather low. However, this happens
when the shock is well outside the galaxy, and our conclusions are
not affected.

The ejecta content inside the galaxy is $M_{\rm ej}=1.7\times
10^5$ \msun after 30 Myr ($<{\cal Z}_{\rm gal}>\sim 1.2\times 10^{-3}$) 
and decreases steadily down to $M_{\rm
ej}=0.77\times 10^5$ \msun after 125 Myr, when $<{\cal Z}_{\rm gal}>\sim
8.3\times 10^{-4}$. At $t=125$ Myr we find $f_{\rm ej}=0.88$.

It is particularly interesting to investigate the X-ray emission for
model BCOND, since now the numerical diffusivity does not affect
the value of $L_X$ and $<{\cal Z}>_X$ (cfr. section 3.4).
In fact, the thermal conduction
naturally broadens the contact surfaces on lengthscales larger than
thickness due to the numerical diffusion. The temporal variation of $L_X$
is shown in Fig. 6. $L_X$ is higher than in model B because of the
emission arising in conduction fronts. The emission averaged abundance
$<{\cal Z}>_X$ ranges between 0.13 and 0.20
for $t\lta 70$ Myr. 
As the energy input stops, the temperatures of the hot phase quickly drops
and after $t\sim 70$ Myr no more X-ray emitting gas is present.
The observable emission averaged temperature of 
the hot gas is $<T>_X\sim 2 \times
10^6$ K for $t\lta 30$ Myr and drops quickly thereafter.

\section{Discussion and Conclusions}

The results presented here qualitatively confirm the conclusions 
of previous investigations on the effect of galactic winds in dwarf galaxies
(e.g. MF).
In general, it is found that the ISM is more robust than expected,
and it is not disrupted even if the total energy input is much greater
than the gas binding energy. In fact, the gas in the optical
region of dwarf galaxies is only temporarily affected by the starburst,
and the galaxy is able to recover a `normal' ISM after a time of the
order of 100 Myr from the starburst event, here assumed to be instantaneous.
Our results agree well with the `moderate form' of galactic wind
dominated evolution of dwarf galaxies described by Skillman (1997).
In Table 2 we summarize the values of the fraction of ISM and metal-rich 
ejecta that is lost by the galaxy.

We find that the evolution of the ISM can be separated in two phases.
The first one corresponds to the energy input period (which lasts 30 Myr
in our models). During this phase the superbubble expands
surrounded by a fragmented and filamentary shell of cold gas.
The hot gas inside the bubble and the cold shell gas are in pressure
equilibrium.
The second phase starts when the energy input stops: the pressure
of the hot bubble, still expanding along the polar direction, 
drops quickly. This causes the inner portion of the shell 
near the equator to
collapse back toward the center, replenishing the galactic region
with cold gas. The collapse is driven mainly by the pressure gradient,
with the gravity being of secondary importance.

\begin{table}
 \caption{Global derived quantities.}
 \label{symbols}
 \begin{tabular}{@{}lccccc}
  Model & $f_{\rm ISM}$ & $f_{\rm ej}$
        & $M_{\rm ISM,gal}$
        & $M_{\rm ej,gal}$ & $<{\cal Z}_{\rm gal} >$ \\
  STD   & 0.058 & 0.46 & $3.4 \times 10^6$  & $5.15 \times 10^3$
        & $1.4 \times 10^{-3}$ \\
  SB1   & $\sim 0$ & 0.003 & $6.5 \times 10^7$  & $2.4 \times 10^4$
        & $3.7 \times 10^{-4}$ \\
  PEXT  & 0.31 & 0.83 & $1.9 \times 10^6$  & $5.8 \times 10^3$
        & $3 \times 10^{-3}$ \\
  B     & 0.48 & 0.77 & $9.3 \times 10^7$  & $ 10^5$
        & $1.1 \times 10^{-3}$ \\
  BCOND & 0.32 & 0.88 & $9.2 \times 10^7$  & $7.6 \times 10^4$
        & $8.3 \times 10^{-4}$ \\
 \end{tabular}
 \medskip

Quantities are calculated at $t=200$ Myr, except for model B 
and BCOND, for which the final time is 106 Myr and 125 Myr, respectively.
$f_{\rm ISM}$ is the ISM ejection efficiency defined in section 3.2.
$f_{\rm ej}$ is the metals ejection efficiency defined in section 3.3.
$<{\cal Z}_{\rm gal} >$ is the average ejecta fraction in the
galactic region (see section 3.3). Masses are given in $M_\odot$.
\end{table}

The replenishment process occurs through inflow of cold gas moving
parallel to the equatorial plane, thus resembling the inflows
considered by Tenorio-Tagle \& Munoz-Tunon (1997). However, the ram pressure
associated to this flow in model STD, representative of all our models, 
is $\approx 10^{-14}$ dyn cm$^{-2}$, five orders of magnitude lower 
than those assumed by Tenorio-Tagle \& Munoz-Tunon (1997).
Evidently, if such massive inflows exist, they must have a different
origin.

We found that the central ISM reaches the critical column density
required for rapid star formation after 0.1 - 1 Gyr from the starbust, 
the exact value depending
on the galactic parameters, when a new starburst may
start. This episodic star formation regime is necessary to account
for the chemical evolution of BCD galaxies, and we have shown here
that it is consistent with the hydrodynamical evolution of the ISM.

Most of the metal-rich material shed by the massive stars 
resides in the hot phase of the ISM, and for powerful starbursts it is easily
lost from the galaxy (Table 2). We estimate that a fraction of $0.5-0.9$ of the
total metal-rich gas is dispersed in the intergalactic medium when
the starburst model SB2 is adopted.
However, for moderate energy input rates (model SB1), only a small fraction
($\lta 10$ \%) becomes formally unbound. In spite of the
smallest $f_{\rm ej}$, model 
SB1 has the lowest $<{\cal Z}_{\rm gal}>$, since the total amount of
ejecta is a factor of 10 lower than the other models.
Most of the ejecta material is
pushed to large distance from the galaxy (several kpc), and its fate
is uncertain, being subject to ram pressure and tidal stripping.
These processes may effectively remove material loosely bound to the
galaxy.

There is some quantitative difference between our findings and
those by MF. The generally lower $f_{\rm ej}$
found in our model is likely to be
the result of our more extended gaseous halo. Their models, with a
sharp truncation of the ISM, are similar to our model PEXT.
The most striking disagreement is between model SB1,
for which we find $f_{\rm ej}=0.003$, and their model
with $M_{\rm gas}=10^8$ \msun and $L_{\rm inp}=10^{39}$ erg s$^{-1}$
which has $f_{\rm ej}=1$. However, as explained in section 4.1, model
SB1 suffers of some numerical extra-cooling; the same model without radiative
losses gives $f_{\rm ej}=0.14$, probably a more realistic value if
thermal conduction is not effective. A comparison of our Fig. 7a-b with 
fig. 2a of MF (panel with the model $M_{\rm g}=10^8$; $L_{38}=10$, 
in their notations) dramatically shows the sensitivity
of the superbubble dynamics (and size) on the ISM distribution.

The cold gas replenishing the central region, from which a successive starburst
may form, has been only slightly polluted by the massive star ejecta,
with $<{\cal Z}> \approx 4\times 10^{-4} - 2\times 10^{-3}$, 
or $<Z_{\rm O}> \approx 2\times 10^{-3} - 10^{-2}$ $Z_{\rm O,\odot}$,
with the assumptions described in section 3.3. 
Thus, many starburst
episodes are necessary to build an average metallicity
$Z\sim 0.25$ $Z_\odot$, as determined for NGC 1569 from observations of
emission lines gas (e.g. Kobulnicky \& Skillman 1997). Many dwarf galaxies,
however, are more metal poor (the abundance of IZw18 is only 0.07 solar), 
and few bursts may be sufficient to
produce the observed metallicity.

The values listed in Table 2 demonstrate how the evolution
of the ISM is not regulated by the ejection efficiency parameters
$f_{\rm ISM}$ and $f_{\rm ej}$ alone. For instance, model PEXT and BCOND
have similar $f_{\rm ISM}$ and $f_{\rm ej}$, but very different ISM and
ejecta masses. Conversely, the gas mass evolution of model SB1 and model B
is comparable, despite of dissimilar ejection efficiencies.

Comparing models STD and B, we found that a dark matter halo
has little {\it direct} influence on the final behaviour of the ISM.
However, dark matter and rotation determine the initial gas
distribution, which is an important parameter for the flow evolution.
For instance, the central region of model STD is refilled with
cold gas more slowly than model B, a difference which reflects the
different initial ISM distribution.

In order to evaluate how the assumption of an instantaneous burst
influences our results, we run an
additional model similar to STD,
but with $L_{\rm inp}=1.128 \times 10^{40}$ erg s$^{-1}$,
$\dot M = 0.009$ M$_\odot$ yr$^{-1}$. The energy and mass sources 
are now active
for 100 Myr, so that the total energy and mass injected are the same as 
in STD. With this model, which in a simple way mimics the effect of
a prolonged starburst, we wish to check the sensitivity of our results
on the assumption of a instantaneous starburst adopted in models illustrated
in the previous sections.
We found that the general dynamics is similar to that of model STD 
and is not described here.
The ejection efficiencies, again calculated at $t=200$ Myr, are
$f_{\rm ISM}=0.055$ and $f_{\rm ej}=0.40$, almost identical to those
found for STD (Table 2). Thus, it appears that the instantaneous starburst
hypothesis does not invalidate our general findings.

We devoted section 3.4 to discuss the complex X-ray emission arising
from starburst galaxies. We warn again that for model BCOND only we
can calculate the X-ray quantities in a strictly consistent way,
the other models having the contact surfaces numerically spread
by the intrinsic diffusion of the hydrocode (although several
physical processes are thought to produce similar effect, cfr. section 3.4). 
In model BCOND
the thermal conduction generates {\it physically} broadened interfaces
between hot and cold gas. We found that the X-ray luminosities 
in the ROSAT band (Fig. 7) are generally less than those estimated by Della
Ceca et al. (1996, 1997) for NGC 1569 and NGC 4449. It is thus suggested
that a mass loading mechanism is at work in these systems (see also
Suchkov et al. 1996).
Moreover, the low abundances found in X-ray studies 
also indicate
that thermal conduction (or some other process that mix cold and hot gas)
is effective.

Contrary to Suchkov et al. (1994) we found that the shocked ISM layer does
not contribute appreciably to the X-ray luminosity for $t\lta 30$ Myr,
a fact indicating that the origin of the X-ray radiation
is model dependent. For $L_{\rm inp}$ appropriate for typical dwarf galaxies
we found that $L_X$ is dominated by the ISM mixed with the shocked ejecta
at the contact discontinuities.

\section*{Acknowledgments}
We are grateful to Luca Ciotti, Laura Greggio, Bill Mathews and Monica Tosi
for interesting discussions. We are indebted to the referee for a number of
thoughtful comments and suggestions.
This work has been partially supported by the Italian Ministry of 
Reserch (MURST)
through a Cofin-98 grant.

\bsp

\label{lastpage}

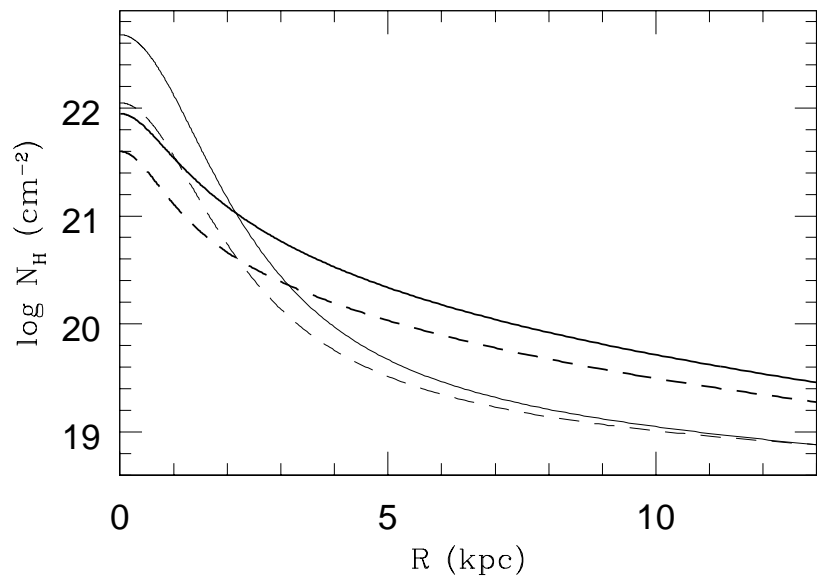
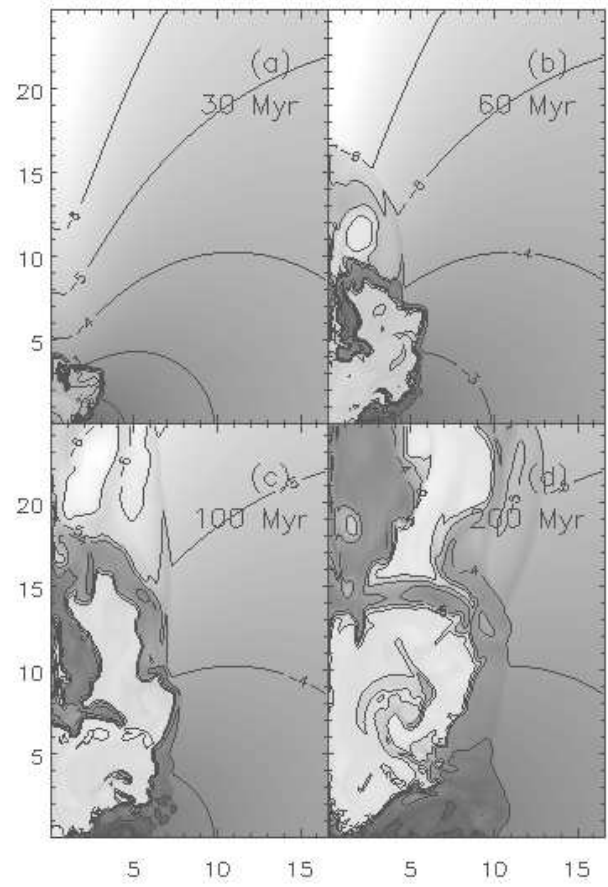
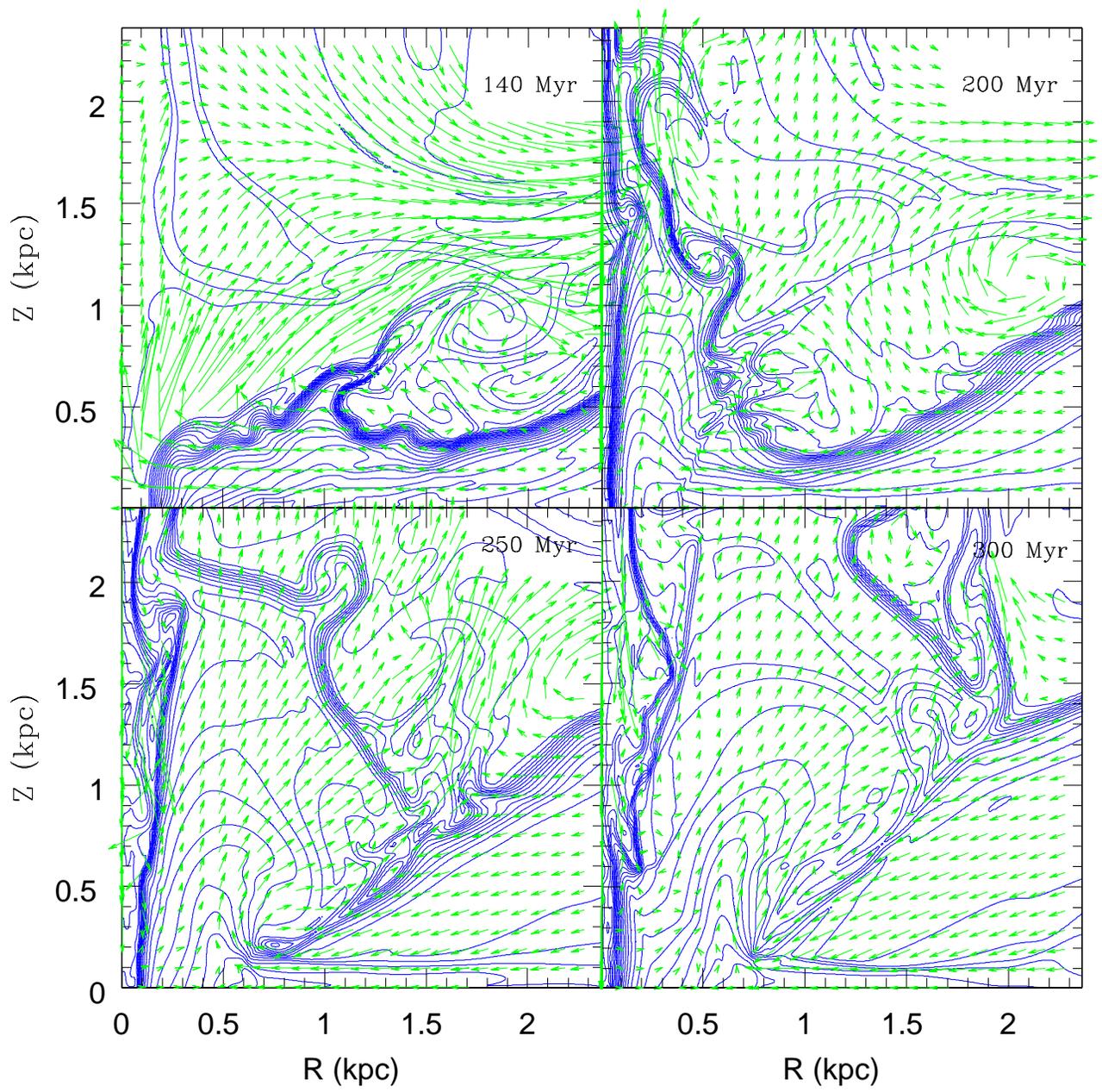
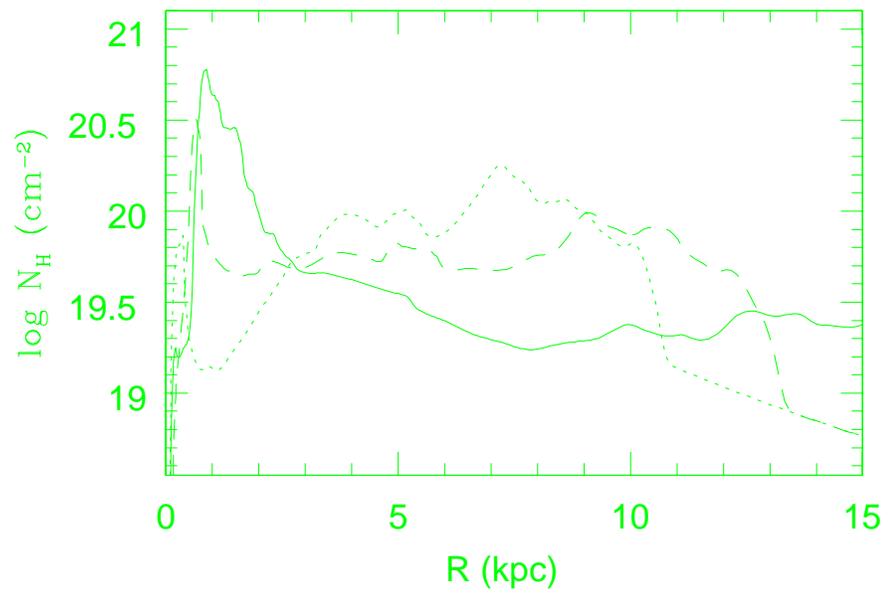
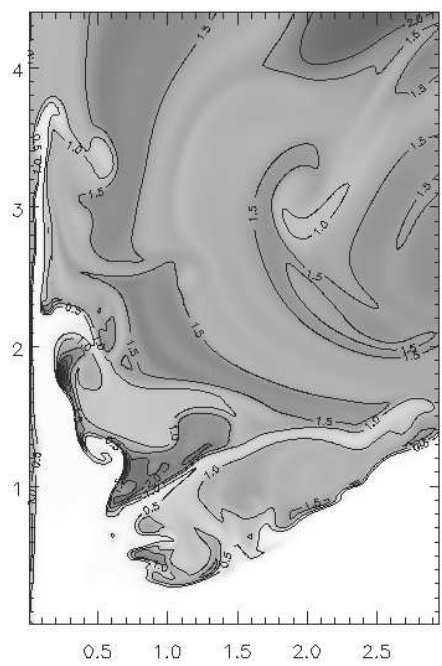
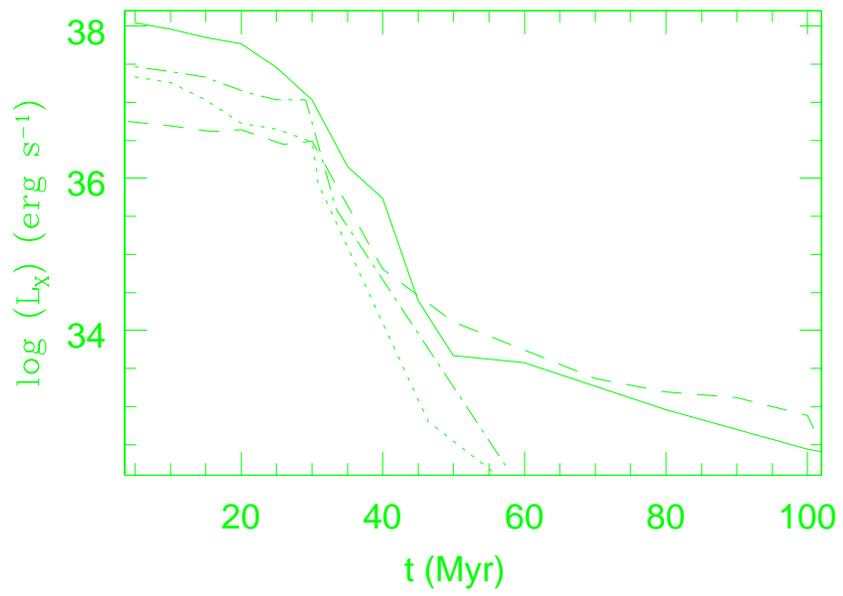
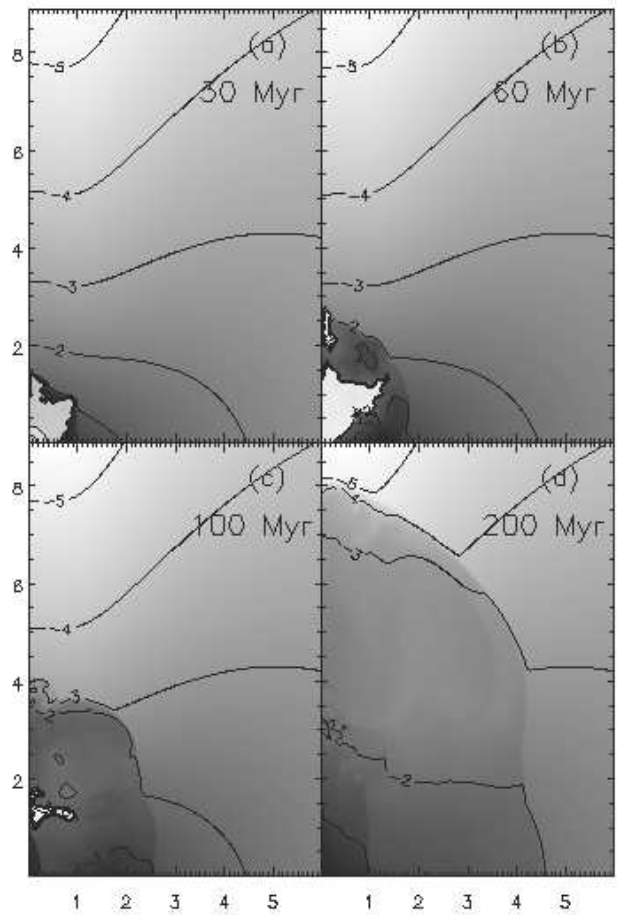
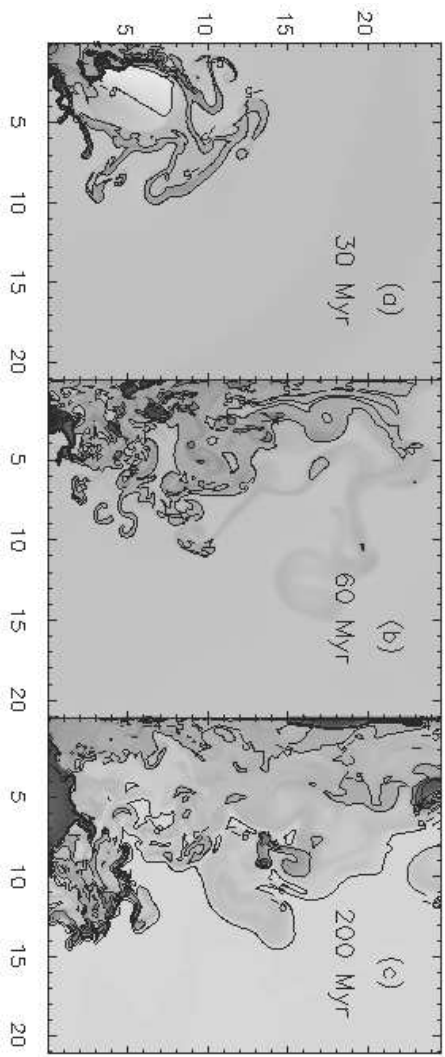
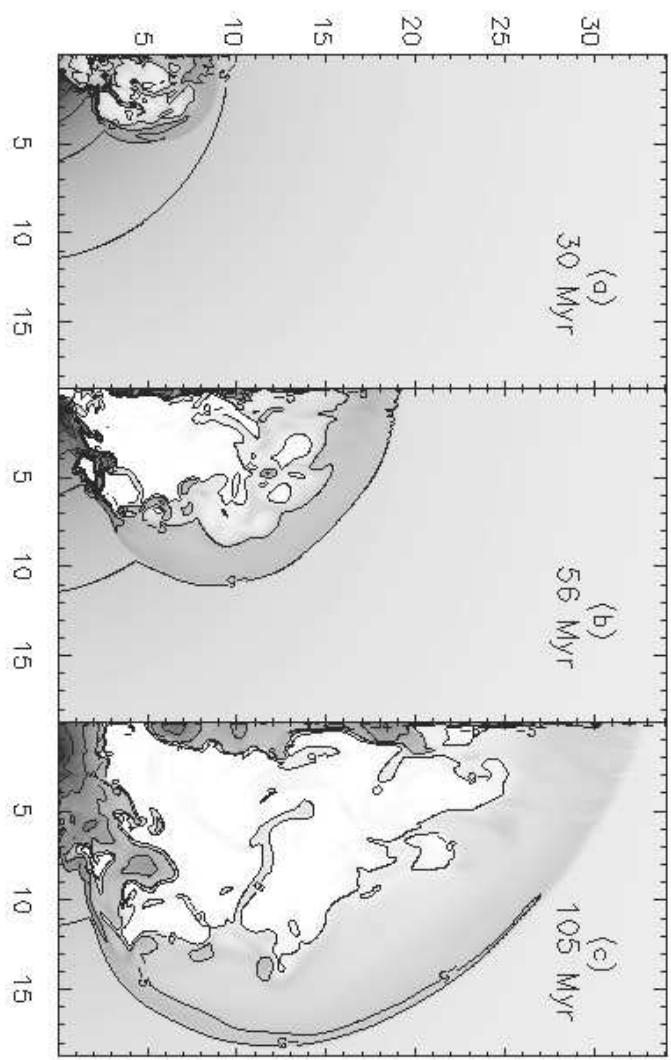
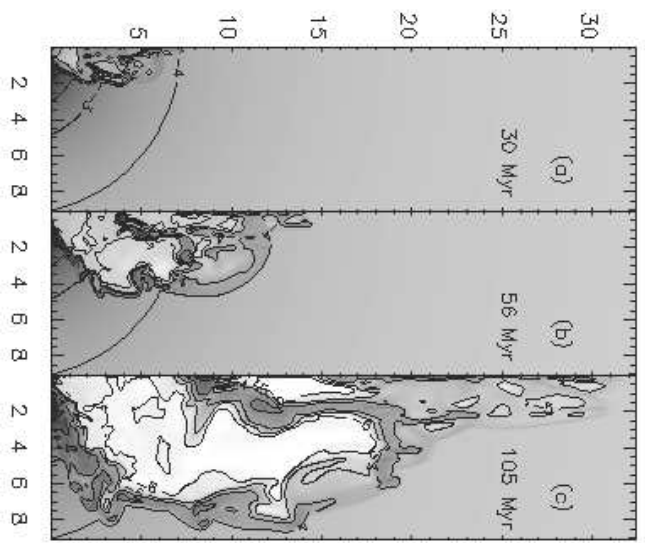
\end{document}